\documentclass[structabstract]{aa}

\usepackage{natbib}
\bibpunct{(}{)}{;}{a}{}{,} 
\usepackage{graphicx}
\usepackage{txfonts}

\newcommand{\pexrav}{{\fontfamily{ptm}\fontseries{m}\fontshape{sc}\selectfont{pexrav}}}

\begin{document}
\title{Origin of the X-ray disc-reflection steep radial emissivity}

\author{J.~Svoboda,$\!^{1}$ M. Dov\v{c}iak,$\!^{2}$ R.~W.~Goosmann,$\!^{3}$ P.~Jethwa,$\!^{1}$ V.~Karas,$\!^{2}$  G.~Miniutti,$\!^{4}$ \and M.~Guainazzi$^{1}$}
\institute{$^{1}$ European Space Astronomy Centre of ESA, P.O. Box 78, Villanueva de la Ca\~{n}ada, 28691 Madrid, Spain \\
$^{2}$ Astronomical Institute, Academy of Sciences, Bo\v{c}n\'{\i}~II~1401, CZ-14131~Prague, Czech~Republic\\
$^{3}$ Observatoire astronomique de Strasbourg, Equipe Hautes Energies, 11 rue de l`Universit\'{e}, F-67000 Strasbourg, France\\
$^{4}$ Centro de Astrobiolog\'{i}a (CSIC--INTA), Dep. de Astrof\'{i}sica; 
ESA, P.O. Box 78, Villanueva de la Ca\~{n}ada, 28691 Madrid, Spain}

\authorrunning{J.~Svoboda et al.}
\titlerunning{Radial emissivity}
\offprints{J.~Svoboda, email: jsvoboda@sciops.esa.int}

\begin{abstract}
{ 
X-ray reflection off the accretion disc surrounding a black hole, 
together with the
associated broad iron K$\alpha$ line, has been widely used to constrain the
innermost accretion-flow geometry and black hole spin. Some 
recent measurements have revealed steep reflection emissivity profiles
in a number of active galactic nuclei and X-ray binaries.
}
{
We explore the physically motivated conditions 
that give rise to the observed steep disc-reflection emissivity
profiles.
}
{
We perform a set of simulations based on the configuration of a possible
future high-resolution X-ray mission.
Computations are carried out for typical X-ray bright Seyfert-1 galaxies.
}
{
We find that steep emissivity profiles with $q\sim 4-5$ (where the
emissivity is $\epsilon (r) \propto r^{-q}$) are produced considering
either i) a lamp--post scenario where a primary compact X--ray source
is located close to the black hole, or ii) the radial dependence of the
disc ionisation state. If both effects are taken into account,
emissivity profiles as steep as $q \sim 7$ can be obtained from X--ray
spectra modelled via conventional reflection models. We also highlight
the role of the reflection angular emissivity: 
the radial emissivity index $q$ is overestimated
when the standard limb--darkening law is used to describe the data.
}
{
Very steep emissivity profiles with $q \geq 7$ are naturally obtained
by applying reflection models 
that take into account radial profile $\xi (r)$ of the disc ionisation
induced by a compact X--ray source located close to the central black
hole.
}

\end{abstract}

\keywords{Black hole physics -- Accretion, accretion discs -- Relativistic processes -- Galaxies: nuclei}

\maketitle

\section{Introduction}\label{intro}

The innermost black-hole accretion discs of active galactic nuclei (AGNs) and black hole binaries
may be revealed through their X-ray radiation released as either the thermal radiation
of the disc \citep{1997ApJ...482L.155Z, 2006csxs.book..157M} 
or the result of the inverse Compton scattering of the thermal photons
on the relativistically moving electrons in the hot corona above the disc
\citep{1976ApJ...204..187S, 1991ApJ...380L..51H}.
While the first is relevant only to stellar-mass black holes where the accretion disc
is heated to very high temperatures (around $10^7$\,K), the second is common
for any accreting black hole over the entire possible mass scale. 

Some fraction of the photons scattered in the corona
reflect off the disc surface before reaching the observer.
This so-called reflection spectrum is characterised by 
a Compton hump at energies of around $20-40$\,keV, 
fluorescent lines of which the iron K$\alpha$ line at $6.4-6.97$\,keV
is the most prominent, and a soft excess below 1\,keV. 
The overall spectral shape depends on the ionisation of the disc surface. 
The spectrum is then smeared by the relativistic effects
including energy shift, aberration, and light-bending. 
The inner disc radius, inclination angle, average ionisation state,
and reflection-emissivity radial profile can in principle be
determined from the resulting X--ray spectral shape. The inner disc
radius plays a particularly important role. As shown e.g. by
\citet{2008ApJ...675.1048R}, the inner-disc reflection radius can be associated
with the innermost stable circular orbit, which only depends on
the black hole spin. Hence, X--ray reflection spectra can be used to
estimate the black hole spin in both AGN and black hole binaries
\citep[see e.g.][and references therein]{2003PhR...377..389R,
2009ApJ...697..900M, 2009ApJ...702.1367B}.

The black hole spin measurements are thus influenced by the geometry
of the disc-illuminating corona and the local properties
of the disc that affect the re-processing and re-emission
of the incident photons. 
%
Current relativistic kernels that are applied to
reflection (and/or iron line) models to include the relativistic
effects on the spectral shape
\citep{1991ApJ...376...90L,
2004ApJS..153..205D, 2004MNRAS.352..353B, 2006ApJ...652.1028B} 
are based on a series of simplified
assumptions and, in particular, they assume a single/broken 
power-law form of the radial reflection emissivity 
and an angular emissivity law defined 
by a simple analytical formula,
most frequently employing a limb-darkening profile \citep{1991ApJ...376...90L}.

In \citet{2009A&A...507....1S}, we pointed out that
the choice of a particular angular emissivity law can 
affect the derived relativistic parameters, including the black hole
spin. In the present work, we extend our analysis to the radial
emissivity law and, in particular, we investigate the physical
conditions under which steep radial emissivity profiles are produced
by considering the effects of (i) primary X-ray source location,
(ii) radial disc ionisation profile, and (iii) angular emissivity law.

The intrinsic disc emissivity is naturally expected 
to decrease with increasing distance,
i.e. the reflection emissivity is
\begin{equation}
 \epsilon(r) = r^{-q},
\end{equation} 
where $q$ is the emissivity index that can be constant
over all radii or a varying quantity.
The thermal dissipation of the disc decreases as $r^{-3}$ 
\citep{1973A&A....24..337S, 1973blho.conf..343N}.
Therefore, the simplest assumption is postulating
the same dependence for the reflection. 
The more energetic photons are injected into the innermost regions,
and so, more intense irradiation of the disc occurs there.
In addition, assuming a point-like X-ray source at height $h$ on the disc axis,
the irradiation of the disc in the absence of any relativistic effect
is proportional to $(r^2 + h^2)^{-3/2}\propto r^{-3}$, as shown
e.g. by \citet{1997ApJ...488..109R}. An emissivity profile with $q=3$
is therefore considered as {\it standard}, while steeper/flatter indices
may need to be explained.


Steep emissivity profiles have been reported from the
analysis of the X-ray spectra of other AGN, such as 
MCG\,-6-30-15 \citep{2002MNRAS.335L...1F, 2004MNRAS.348.1415V, 2007PASJ...59S.315M},
1H0707-495 \citep{2009Natur.459..540F, 2010MNRAS.401.2419Z, 2011MNRAS.414.1269W, 2012MNRAS.422.1914D},
and IRAS\,13224-3809 \citep{2010MNRAS.406.2591P},
as well as black hole binaries, such as XTE\,J1650-500 \citep{2004MNRAS.351..466M}, 
GX\,339-4 \citep{2007ARA&A..45..441M}, and
Cyg\,X-1 \citep{2012arXiv1204.5854F}.
The measured indices reach values up to $q \approx 7$.

To provide a physical picture of the steep radial emissivity in MCG\,-6-30-15,
\citet{2001MNRAS.328L..27W} invoke strong magnetic stresses that should act 
in the innermost region of the system. This should correspond to the enhanced dissipation 
of a considerable amount of energy in the accretion disc at small radii. 
If the magnetic field lines thread the black hole horizon,
the dissipation could be triggered by the magnetic extraction 
of the black hole rotational energy, perhaps via the Blandford-Znajek effect 
\citep{1977MNRAS.179..433B}, but it could be also supplemented 
by a rather efficient slowing of the rotation, as also seen in recent GRMHD simulations
\citep[e.g.][]{2010MNRAS.408..752P}. 
The efficiency of the competing processes still needs to be assessed.

\citet{2000MNRAS.312..817M} 
examined whether the required steep emissivity law as well
as the predicted equivalent width of the cold reflection line of iron 
and the Compton reflection component can be reproduced in a phenomenological (lamp-post) model 
where the X-ray illuminating source is located on the common symmetry 
axis of the black hole and the equatorial accretion disc. They 
suggested that the radial emissivity function of the reflection component 
steepens when the height parameter of the primary irradiation source decreases. 
The enhanced anisotropy of the primary X-rays was identified 
as a likely agent acting in this process.
The emissivity in the XMM-Newton spectrum of MCG\,-6-30-15 was successfully
reproduced by adopting this lamp-post geometry \citep{2002A&A...383L..23M,
2003MNRAS.344L..22M, 2008MNRAS.386..759N}.

%

To explain the steep radial emissivity, 
we explore several simple test models 
and analyse them with the simulated data.
The paper is organised as follows.
Sect.~\ref{data} describes the data generation.
The effects of the lamp-post geometry of the corona,
the angular directionality, 
and the radially stratified ionisation on the radial emissivity are discussed
in Sects.~\ref{section_lamp_post}, ~\ref{section_angular}, and ~\ref{section_radion}, respectively.
An example of the combined effect is presented in Sect.~\ref{section_combi}.
The obtained results are discussed in Sect.~\ref{discussion}
and summarised in Sect.~\ref{conclusions}.

\section{Data simulation}
\label{data}

Our goal is to build a series of physically motivated reflection
models and to examine the radial emissivity law that would be
inferred from deep, high-quality X-ray spectral observations of AGN
and/or black hole X-ray binaries with either current or future X-ray
facilities. We build theoretical reflection spectral models that are
then put into X-ray spectral simulations to produce high quality X-ray
data. The simulated data are then modelled with the currently
available disc-reflection models and standard spectral tools. 
To evaluate the possibility of measuring the simulated effects
with future X-ray missions, we adopted 
a preliminary response matrix for the
Athena X--ray mission \citep{2011xru..conf...22N}. 
This ensures that our results are not
driven/limited by the signal--to--noise that can be reached with
current X-ray observatories.

%
We performed our spectral analysis within the 2-10\,keV energy range
where one of the most prominent reflection feature, the iron K$\alpha$ line, occurs.
The preliminary response matrix,\footnote{ftp://ftp.rssd.esa.int/pub/athena/09052011\_Responses}
was re-binned by a factor of ten between channels 2700 and 8800
(2-10\,keV). The other channels were not used. 
This provided us with sufficient spectral resolution and increase in the computational speed.
We used Xspec \citep{1996ASPC..101...17A}, version 12.6.0ab for the spectral fitting. 
We used the most recent version of the KY code \citep{2004ApJS..153..205D},
which includes the lamp-post geometry (Dov\v{c}iak et al., in prep.).
The flux of the model was chosen to be similar to that of bright Seyfert
galaxies, i.e. $\approx 3 \times 10^{-11}$\,ergs\,cm$^{-2}$\,s$^{-1}$ \citep{2007MNRAS.382..194N}.
The simulated observation time was 100\,ks,
which is comparable to the average exposure time of AGNs
observed with the current X-ray satellites.


\section{Lamp-post scheme}
\label{section_lamp_post}

We first investigate how the radial emissivity depends 
on the geometry of the corona. If the corona is 
localised the illumination of the disc
decreases with the growing distance from the source
in a way determined by the position of the corona
and by the gravitational field of the central black hole. 
The configuration when the corona is very compact 
and located just above the black hole, known also as
the lamp-post scheme, was studied as a simple 
disc-corona scenario by \citet{1991A&A...247...25M}
and \citet{1991MNRAS.249..352G}.
In a physical picture, the source above the black hole can be imagined, 
e.g., as the base of a jet.

In this scenario, the irradiation far from the source 
radially decreases as $r^{-3}$. 
In the central region, the relativistic effects
influence the disc illumination, thus shape 
the reflection spectra of black hole accretion discs
\citep{2004MNRAS.349.1435M}.
As a result, the different
parts of the disc are irradiated with different intensities,
making the emissivity profiles in reflection models
distinct from the standard value of $q=3$.
If the height of the source is sufficiently close to the black hole
event horizon, light bending implies higher irradiation of the innermost
region compared to the outer parts of the disc.

The exact profile of the radial emissivity depends
on the geometrical properties of the source.
Different cases of axial, orbiting,
jet, and extended sources
were studied by \citet{2012arXiv1205.3179W}.
The steepest profiles were obtained for point-like sources
at small heights along the vertical axis.
In the most extreme scenario, when the source 
is moving towards the black hole, the Doppler 
boosting might increase this effect. However,
there is no observational evidence for such an inflow
of the matter perpendicular to the disc plane, while
outflows in the form of a jet or a disc wind
are observed in many sources
\citep[e.g.][]{2003MNRAS.345.1057M,blustin05,2007ApJ...659.1022K,2010A&A...521A..57T}.

\begin{figure*}
\includegraphics[width=0.5\linewidth]{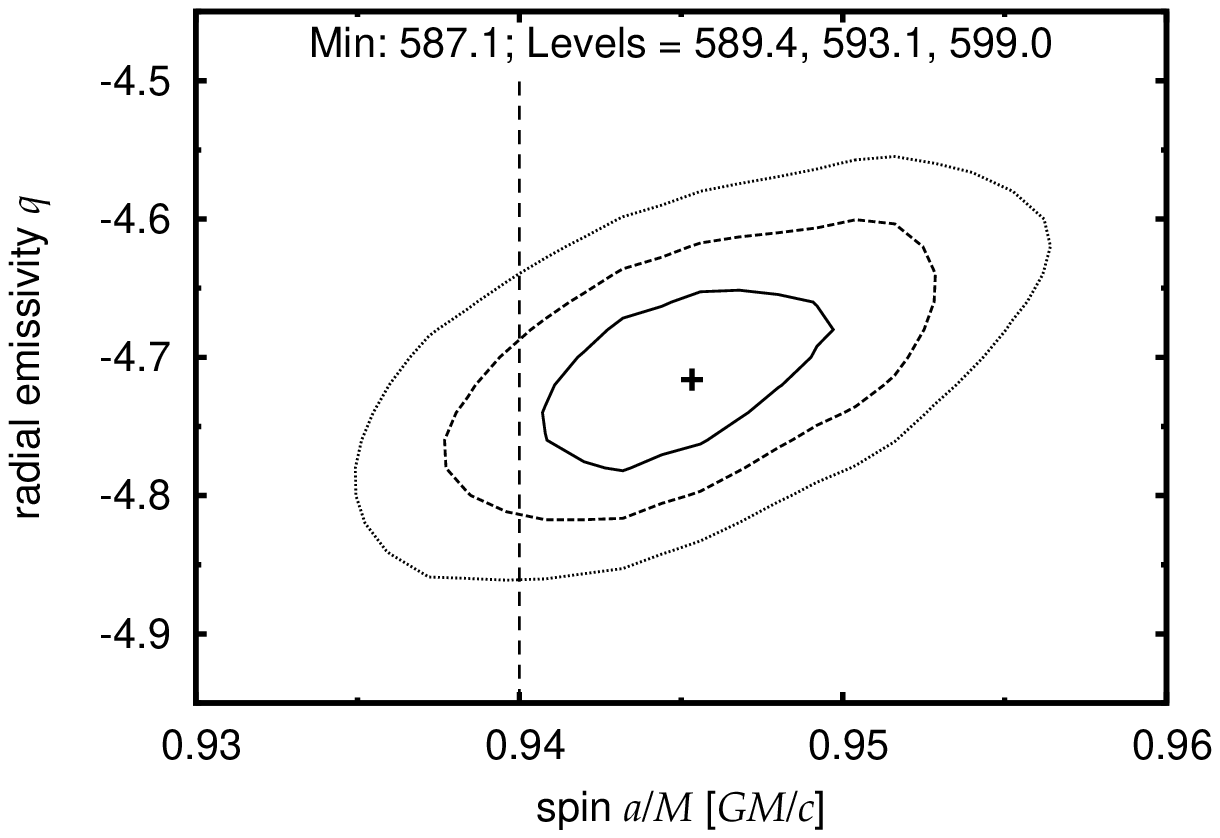} \hfill
\includegraphics[width=0.5\linewidth]{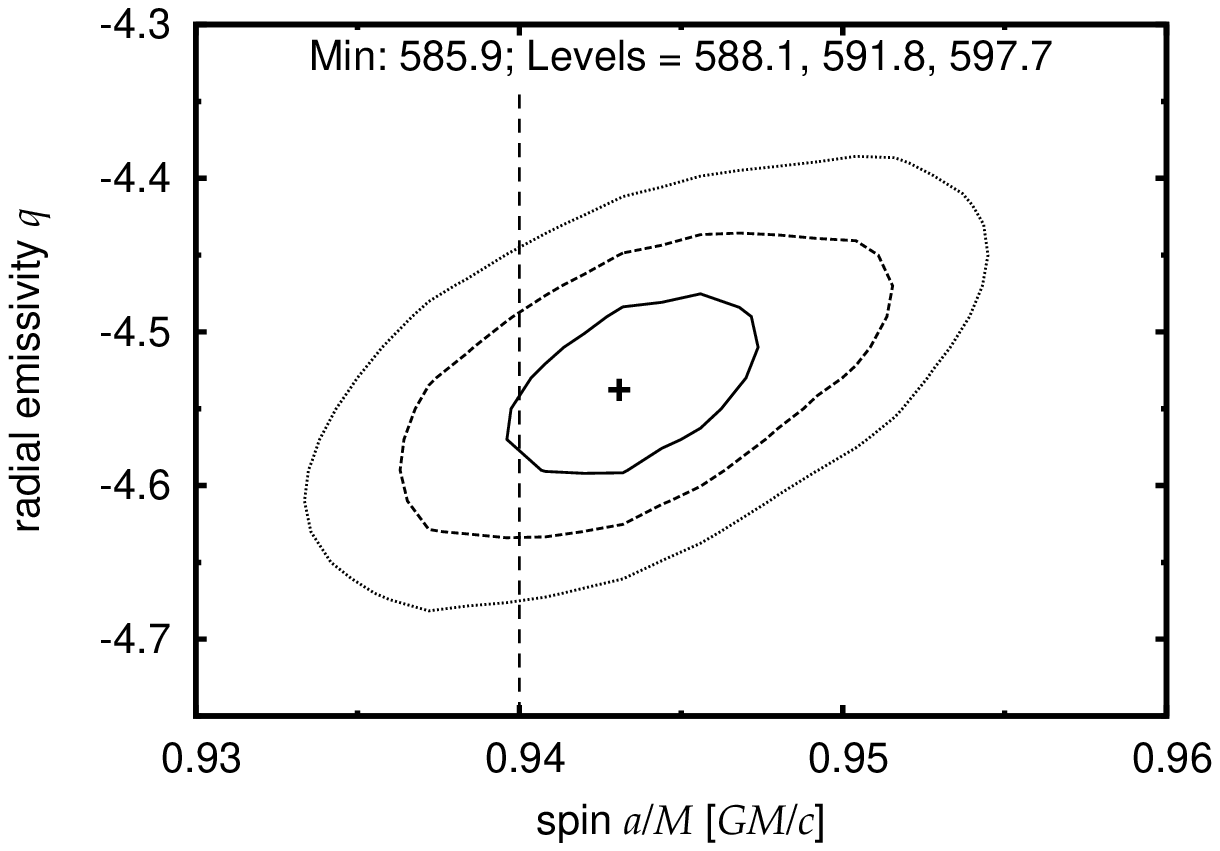}
\includegraphics[width=0.5\linewidth]{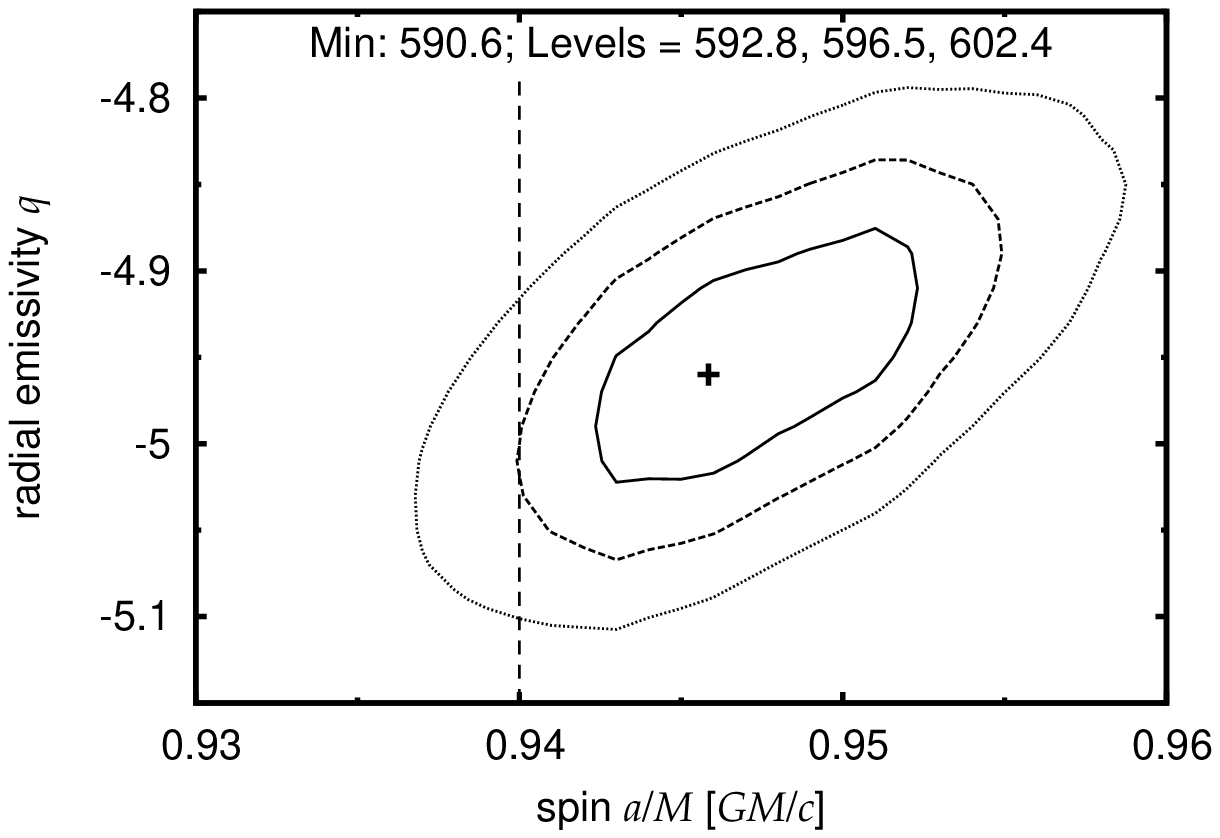} \hfill
\includegraphics[width=0.5\linewidth]{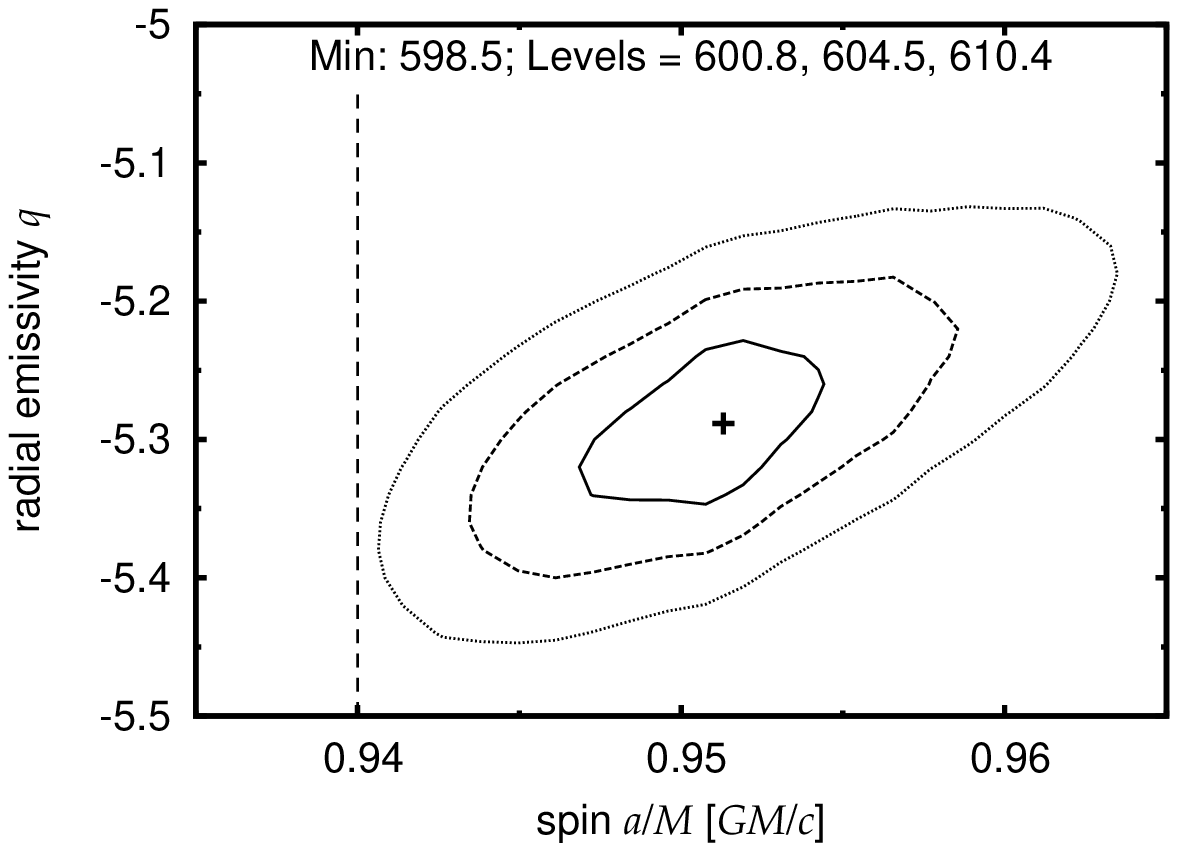}
\caption{Contour plots of the spin $a$ and the radial emissivity parameter $q$. The data were generated with the lamp-post model
with the height $h=1.5\,r_{\rm g}$. The default value of the spin was $a=0.94$, which is indicated by a dashed line in the graph.
Different prescriptions for the angular emissivity were used:
{\textbf{Top left:}} Angular emissivity from numerical calculations.
{\textbf{Top right:}} Limb brightening.
{\textbf{Bottom left:}} Isotropic.
{\textbf{Bottom right:}} Limb darkening.
The $\chi^2$ values corresponding to the best fit (indicated by a small cross) 
and to the $1\sigma, 2\sigma$, and $3\sigma$ levels are shown in the text legend.}
\label{kyfix_contaq_a094i30h015}
\end{figure*}

\begin{table*}
\begin{center}
\caption{\label{table_angdir} Inner radial emissivity index $q_{\rm in}$
and the break radius $r_{\rm b}$ inferred for a different height and directionality
in the lamp-post model.}
\begin{tabular}{c|cc|cc|cc|cc} 
\multicolumn{9}{c}{$a=0.94$}\\
\hline
\hline
\rule[-0.5em]{0pt}{1.6em}
 & \multicolumn{2}{c|}{numerical} & \multicolumn{2}{c|}{limb brightening} & \multicolumn{2}{c|}{isotropic} & \multicolumn{2}{c}{limb darkening}\\
\hline
\rule[-0.5em]{0pt}{1.6em}
$h [r_g]$  & $q_{\rm in}$ & $r_{\rm b}$ & $q_{\rm in}$ & $r_{\rm b}$ & $q_{\rm in}$ & $r_{\rm b}$ & $q_{\rm in}$ & $r_{\rm b}$ \\
\hline
\rule[-0.5em]{0pt}{1.6em}
1.5 &	$4.72^{+0.07}_{-0.08}$	&	$6.1^{+0.2}_{-0.2}$	&	$4.54^{+0.06}_{-0.07}$	&	$6.4^{+0.2}_{-0.2}$	&
	$4.96^{+0.07}_{-0.08}$	&	$6.0^{+0.3}_{-0.2}$	&	$5.29^{+0.06}_{-0.07}$	&	$5.7^{+0.2}_{-0.1}$	\\
\rule[-0.5em]{0pt}{1.6em}
3.0 &	$3.3^{+0.1}_{-0.1}$	&	$6.3^{+1.9}_{-1.1}$	&	$3.2^{+0.3}_{-0.2}$	&	$8.0^{+3.4}_{-2.6}$	&
	$3.2^{+0.1}_{-0.1}$	&	$15^{+10}_{-3}$ 	&	$3.3^{+0.1}_{-0.1}$	&	$20^{+70}_{-5}$	\\
\rule[-0.5em]{0pt}{1.6em}
10 &	$1.3^{+0.1}_{-0.2}$	&	$16^{+1}_{-1}$	&	$2.3^{+0.1}_{-0.1}$	&	$55^{+9}_{-10}$	&
	$2.3^{+0.2}_{-0.1}$	&	$48^{+5}_{-7}$ 	&	$2.5^{+0.1}_{-0.1}$	&	$60^{+15}_{-15}$	\\
\hline
\end{tabular}
\end{center}
\end{table*}

\begin{figure*}
\includegraphics[width=0.5\linewidth]{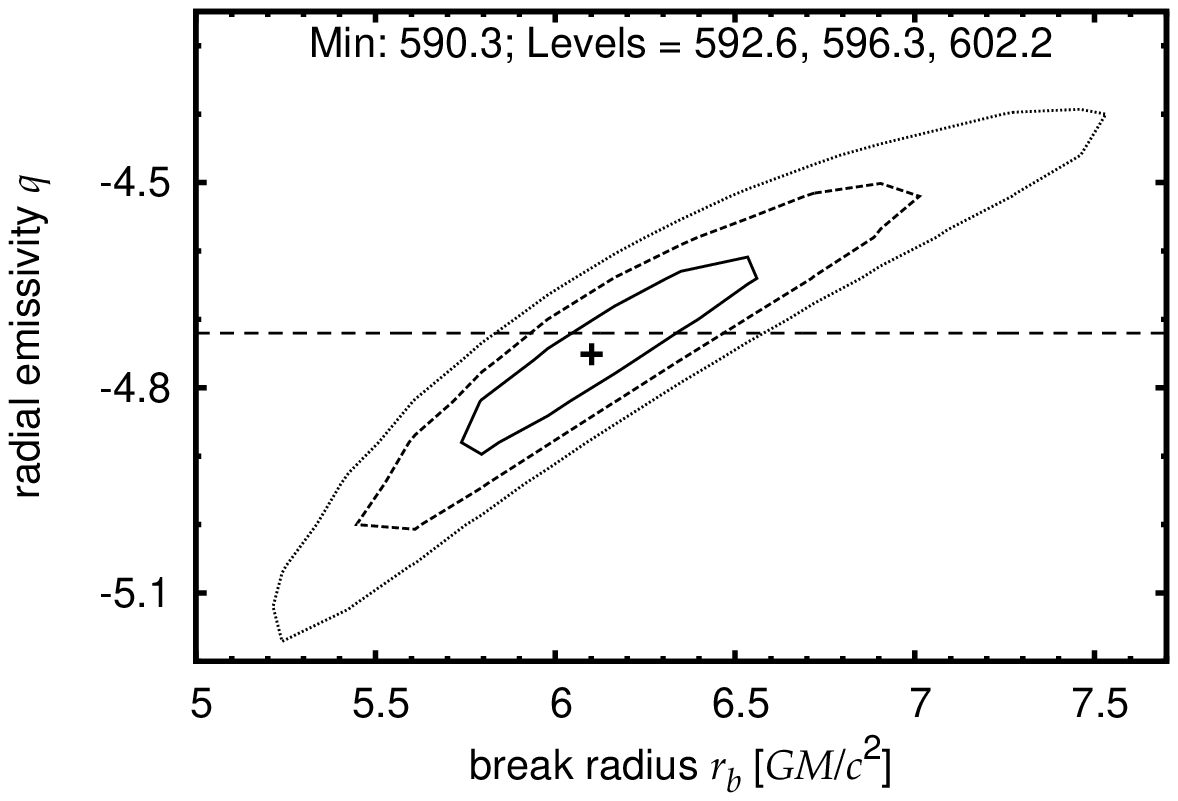} \hfill
\includegraphics[width=0.5\linewidth]{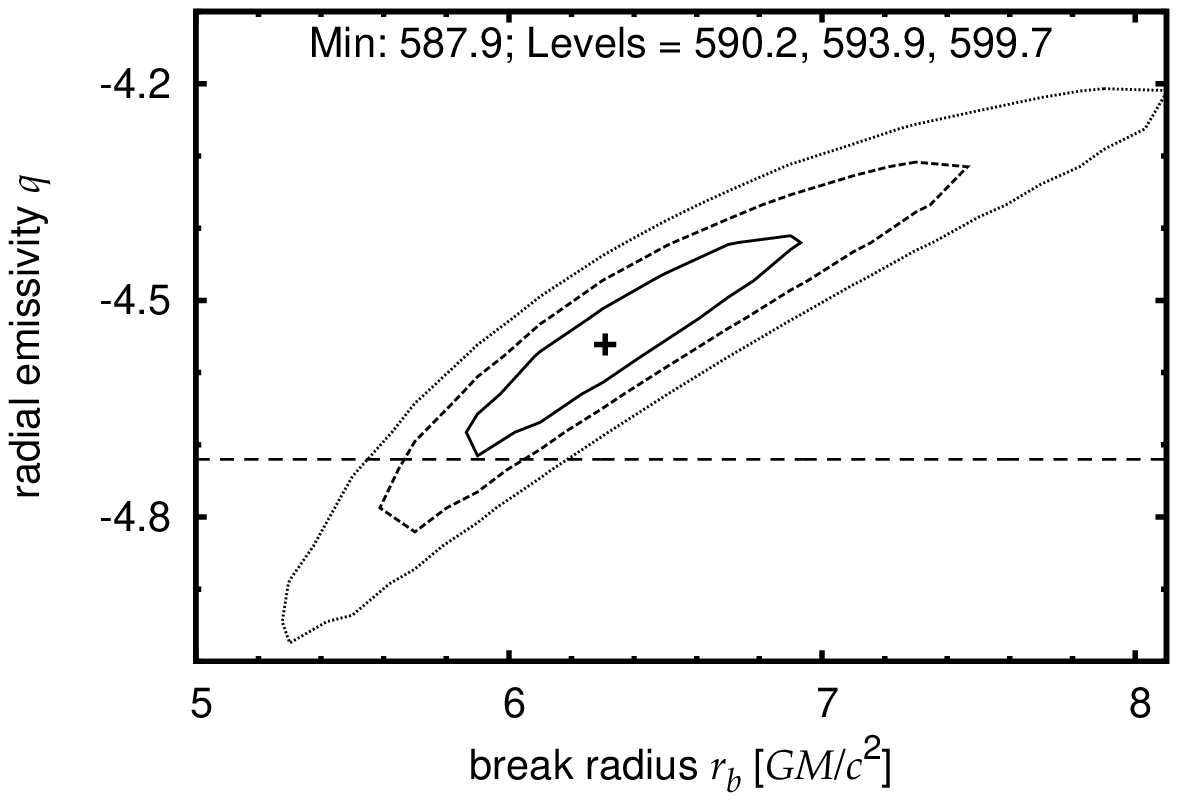}
\includegraphics[width=0.5\linewidth]{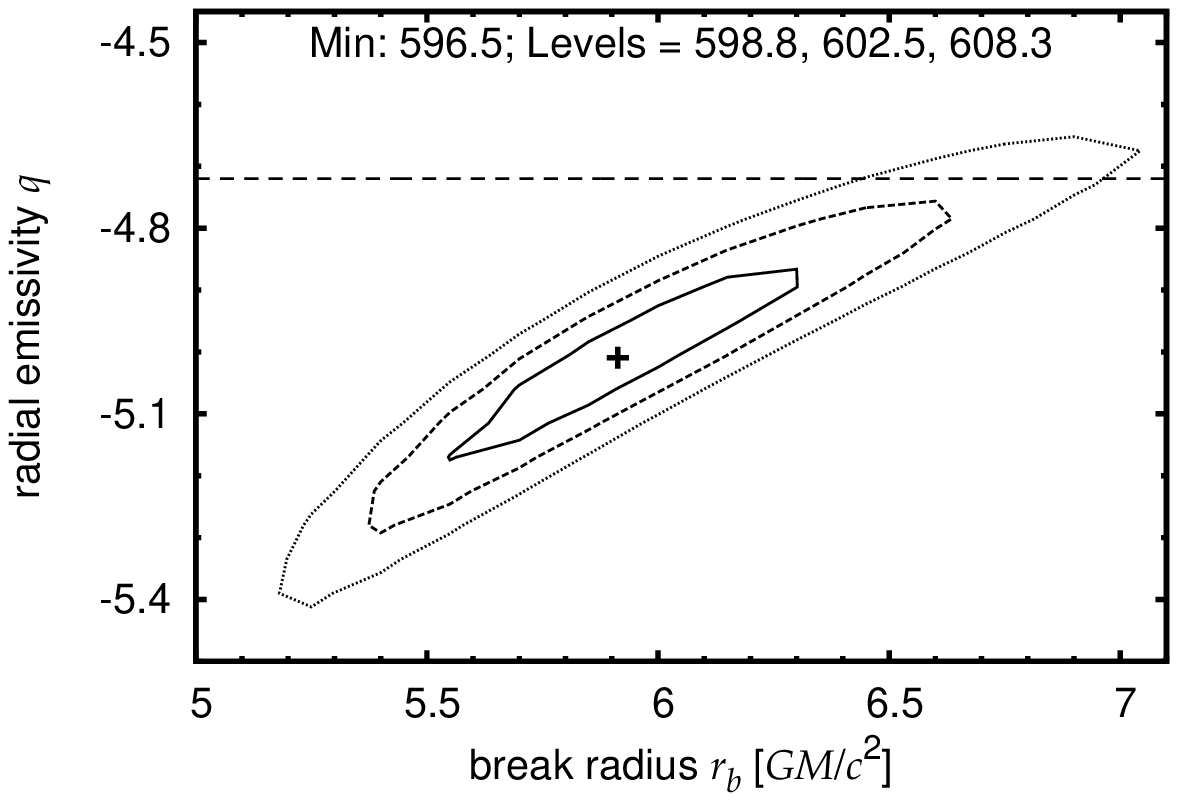} \hfill
\includegraphics[width=0.5\linewidth]{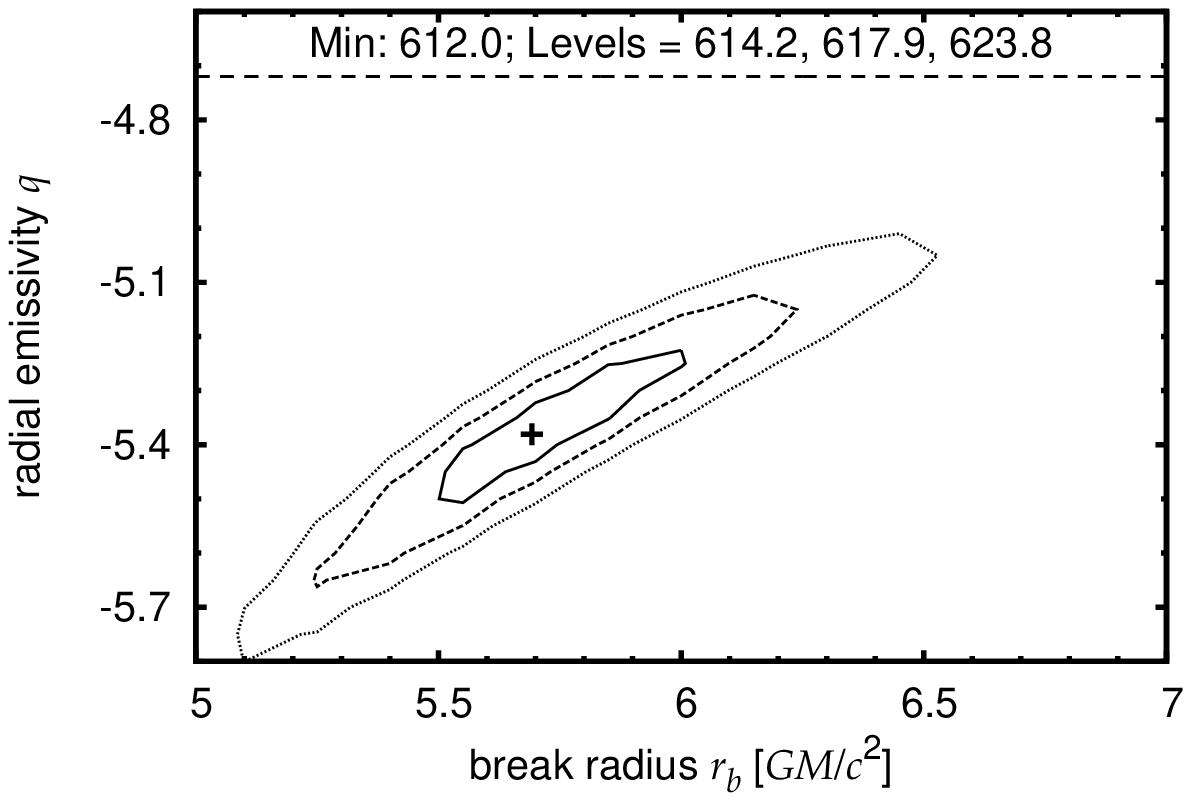}
\caption{Contour plots of the radial emissivity and the break radius 
for the height $h=1.5\,r_{\rm g}$. The legend is the same as in Fig.~\ref{kyfix_contaq_a094i30h015}.
A horizontal dashed line representing the 
best-fit value from Table~\ref{table_angdir}, $q=4.72$, is in the graph only for comparison.}
\label{kyfix_contrbq_a094i30h015}
\end{figure*}

In the following, we consider the stationary lamp-post source
and investigate the radial emissivity profile of the disc
reflection radiation for different heights of the source.
The physical set-up of the model is a combination 
of the general-relativistic lamp-post scheme 
for an X-ray illuminated accretion disc near 
a rotating black hole \citep{2000MNRAS.312..817M} 
and a self-consistent Monte Carlo scheme 
for the X-ray reprocessing within the disc environment \citep{2006AN....327..977G}.
We used local (re-) emission tables that were computed by the radiative transfer code NOAR \citep[][Sect. 5]{2000A&A...357..823D}
for the case of ``cold'' reflection (i.e. for neutral or weakly ionised matter). Photo-absorption,
Compton scattering, and the fluorescent emission of the
iron K line are considered.
A stratified plane-parallel atmosphere
irradiated by a power law with the photon index $\Gamma = 1.9$
is assumed. A large number of primary photons were sampled 
in the 2 -- 300\,keV energy range to ensure sufficiently high quality statistics. 
At all local emission angles, the Poissonian noise level 
is much smaller than any relevant spectral feature.
The computations were done for various
incident local emission angles, both polar and azimuthal.

In the present paper, we employ a new implementation 
of the lamp-post scenario that is based on the KY package \citep[][a more detailed description of the new model is presented in the parallel paper Dov\v{c}iak et al., in prep.]{2004ApJS..153..205D}. 
All the relativistic effects on the photon energies
and trajectories, such as aberration
and light bending, are incorporated in the model.
Here, we study whether it is possible to approximate the radial emissivity
of the lamp-post model by a simplified profile in the form of a broken
power-law, as usually done in current modelling of the data.

For this purpose, we generated data as described in Sect.~\ref{data}
for a power-law model with 
photon index $\Gamma=1.9$ and the relativistic line
in the lamp-post geometry described above.
The spin parameter was set to $a=0.94$
(in the geometrised units where $c=G=M=1$).
The other parameters are the inclination angle of the disc $i=30$\,deg,
which is a typical value for the inclination
of Seyfert~1 galaxies \citep{2010A&A...524A..50D}, 
the inner radius coinciding with the marginally stable orbit
($r_{\rm in}=r_{\rm ms}$)
and the outer radius corresponding to $r_{\rm out}= 400\,r_{\rm g}$,
where $r_{\rm g} \equiv \frac{GM}{c^2}$ is the gravitational radius.
We considered three cases of the source height,
$h=1.5$, $3$, and $10\,r_{\rm g}$. 

We then fitted the radial emissivity profile resulting from the
simulations with a
broken power-law model in which the emissivity index is $q_{\rm in}$
within a break radius $r_{\rm b}$ and $q_{\rm out} = 3$ at larger
distances. As expected, we found that $q_{\rm in}$ increases as the
source height $h$ decreases, while $r_{\rm b}$ decreases. Steep
emissivity profiles in the inner regions (i.e. $q_{\rm in}\geq 3$) were
only obtained for $h\lesssim 3~r_g$, as summarised in Table~\ref{table_angdir}
in the column labelled ``numerical''.


\section{Interplay between the radial and angular emissivity profiles}
\label{section_angular}

When fitting the data, the local intensity of the re-processed radiation emitted from
the disc is often assumed to be divided into two separate parts --
the radial and angular dependence. The latter one characterises
the emission directionality.
However, owing to the large rotational velocity of the disc
and the strong gravity near the black hole, the photons
that reach the observer are emitted under different angles
at different locations. Therefore, the angular part
of the emissivity depends on radius as well, and the above
separation is invalid. The relativistic effects, namely aberration
and light bending, cause the emission angle in the innermost
region to always be very high (almost 90 degrees with respect to the
disc normal) -- see Appendix C in \citet{2004astro.ph.11605D}, 
or Fig.~3 in \citet{2009A&A...507....1S}. Although it is not
an axisymmetric problem, the almost radial decrease in the emission
angle is readily apparent, which invokes the link between the radial
and angular emissivity \citep{2004MNRAS.352..353B, 2009A&A...507....1S}.

To show the impact of the directionality on the radial emissivity profile,
we considered the cases of
\begin{enumerate}
  \item Our numerical computations,
  \item Limb brightening $I(\mu_{\rm e}) \approx \ln(1+\mu^{-1}_{\rm e})$ \citep{1993ApJ...413..680H},
  \item Isotropic emissivity,
  \item Limb darkening $I(\mu_{\rm e}) \approx 1 + 2.06\mu_{\rm e}$ \citep{1991ApJ...376...90L}, 
\end{enumerate}
where $\mu_{\rm e}$ is the cosine of the local emission angle.
The simulated data were created with our numerical model
of the directionality integrated over all incident angles.
The {\sc{NOAR}} code \citep{2000A&A...357..823D}
was employed in the calculations.
Free-free absorption, the recombination
continua of hydrogen- and helium- like ions,
and both direct and inverse Compton scattering were taken into account.

Figures~\ref{kyfix_contaq_a094i30h015} and \ref{kyfix_contrbq_a094i30h015}
show contour plots of the radial emissivity index
and the spin, and the break radius, respectively, 
for different prescriptions of the angular directionality.
The lamp-post model with the source height at $h=1.5\,r_{\rm g}$
was used in the data simulation, while
the fitting model employed a broken power-law emissivity. 
In the contour calculations, only the two parameters of interest 
were allowed to vary. The others were fixed to their default
or best-fit (in the case of break radius) values.
The best-fit parameters are summarised in Table~\ref{table_angdir}.

The worst fit was obtained for the model with limb darkening
for which significantly steeper values of the radial emissivity were required. 
This emissivity law is widely used in the reflection models. However,
it somewhat contradicts several models of X-ray illuminated disc atmospheres
\citep{1994MNRAS.267..743G, 1994MNRAS.266..653Z, 2006AN....327..977G, 2011A&A...527A..47R}.
Its application causes an appreciable underestimation of the innermost flux.
The fit therefore requires an artificially steep value 
of the radial emissivity index
in order to compensate for this loss of counts from
the central region where the emission angle is grazing.



\begin{figure}
\includegraphics[width=0.5\textwidth]{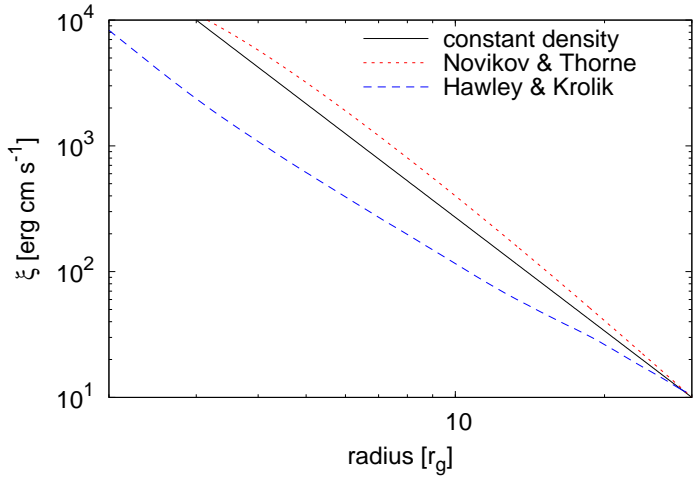}
 \caption{Radial dependence of the ionisation parameter for different
prescriptions of the density: constant (black, solid), Novikov-Thorne profile (red, dotted),
and from the GRMHD simulations by \citet{2006ApJ...641..103H} (blue, dashed). 
The assumed irradiation is $F_{\rm inc} \propto r^{-3}$.
The considered spin value is $a=0.99$. }
 \label{xi_profile}
\end{figure}




\section{Radially structured ionisation of the disc}
\label{section_radion}


The interplay between the radial and angular emissivity shows
that the steep radial emissivity in the observational data 
might be caused by an invalid model assumption.
Yet, there is another frequently used assumption in the reflection scenario
that can contribute to this effect as well -- the constant ionisation 
over the whole surface of the disc. 
The intensity of the disc irradiation, whether approximated
by a (broken) power-law decrease or a lamp-post illumination
in curved space-time, decreases with radius.
Therefore, the ionisation of the disc surface may respond accordingly
\citep[see e.g.][]{1993MNRAS.262..179M}.

The importance of the photoionisation of the disc surface
in AGNs was studied by several authors
\citep{1994ApJ...437..597Z, 2000ApJ...537..833N, 2001MNRAS.327...10B,
2011ApJ...734..112B}.
\citet{2003MNRAS.342..239B} interpreted the X-ray reflection spectrum
of MCG\,-6-30-15 as being composed of 
a highly ionised reflection from the innermost region
and a cold one from the outermost accretion disc.
The radially dependent ionisation in the existing data 
was also discussed by \citet{2011MNRAS.413L..61Z}.
The photoionisation was also suggested
as a possible explanation of the non-detection of the spectral
imprints of the relativistically smeared reflection in some sources
\citep{2004MNRAS.352..205R, 2010A&A...512A..62S, 2011MNRAS.416..629B, 2012ApJ...744...13B}. 

In general, the ionisation of the disc surface  
depends on several other physical quantities such as the density, 
vertical structure, and thermal heating 
\citep[see e.g.][and references therein]{2001ApJ...546..406N, 2002MNRAS.332..799R, 2007A&A...475..155G}.
In particular, \citet{2000ApJ...537..833N}
showed that considering hydrostatic and ionisation balance
and radiative transfer in a plane-parallel geometry
resulted in a cold core within the accretion disc 
and highly ionised outer layers.
A detailed description of the disc ionisation is beyond the scope of this paper.
Here, we simply assume that the radial dependence of the ionisation may be
relevant, as a natural consequence of the radial dependence of the disc 
illumination by the primary radiation. 

If the photoionisation is the dominating factor determining the state of the plasma the ionisation parameter can be defined as \citep[see e.g.][]{1993MNRAS.261...74R}
\begin{equation}
 \xi=\frac{4 \pi F_{\rm inc}}{n_{\rm H}},
\label{xi}
\end{equation}
where $F_{\rm inc}$ is the incident (irradiation) flux and $n_{\rm H}$ is the number
density of hydrogen. Solar abundances are assumed for the other elements.
Typically, X-ray spectra are fitted using only one reflection component, i.e. assuming 
$\xi$ constant over the whole disc. However, in general, eq. \ref{xi}
may be re-written as
\begin{equation}
 \xi=\frac{4 \pi F_{\rm inc}(r)}{n_{\rm H}(r)},
\label{xi_r}
\end{equation}
with the radial dependence of the irradiating flux and the density.

Figure~\ref{xi_profile} shows the radial dependence of the ionisation 
for different density profiles: constant disc density, Novikov-Thorne \citep{1973blho.conf..343N},
and a profile from general relativistic magneto-hydrodynamic simulations
by \citet{2006ApJ...641..103H} for isotropic irradiation decreasing like
$F_{\rm inc} \propto r^{-3}$.
In all cases, the ionisation decreases with the radius 
according to a decreasing strength of the irradiation. 
Although an unknown density profile introduces an uncertainty
in the radial ionisation description,
its effect is, however, not as strong a function
of the radius as the irradiation profile itself.
We therefore assumed a constant
density over the whole disc surface in our more detailed analysis.

To describe the ionisation profile, 
a specific ionisation parameter has to be set at a given radius.
In general, this normalisation depends on the luminosity and the distance
of the irradiating source, in addition to the density of the disc. It is therefore
different for different kinds of sources, and moreover, depends on
the variable accretion rate.
The choice of the normalisation factor is thus somewhat arbitrary
for our purposes.
Here we assumed that the ionisation of the disc is 
$\xi=10$\,ergs\,cm\,s$^{-1}$ at $30\,r_{\rm g}$
since the neutral iron line is observed in most spectra of AGNs.

\begin{figure*}
\centering
\includegraphics[width=0.8\textwidth]{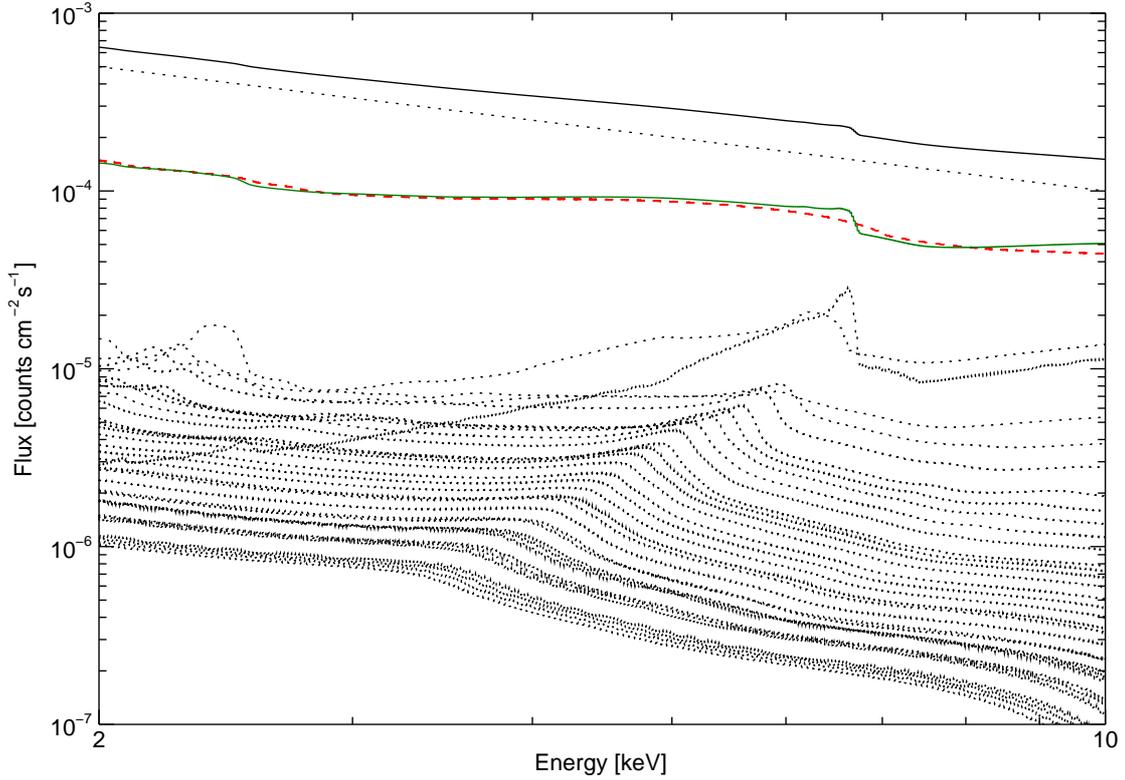}
\caption{Plot of the `complex' model.
The top solid line shows the total model flux, while
the dotted lines are its components. 
The sum of all reflection components is shown
by the green solid line. The red dashed line represents 
the best-fit reflection model assuming single ionisation.
See the main text for more details.
}
\label{model_complex}
\end{figure*}

\begin{figure*}
\includegraphics[width=0.5\textwidth]{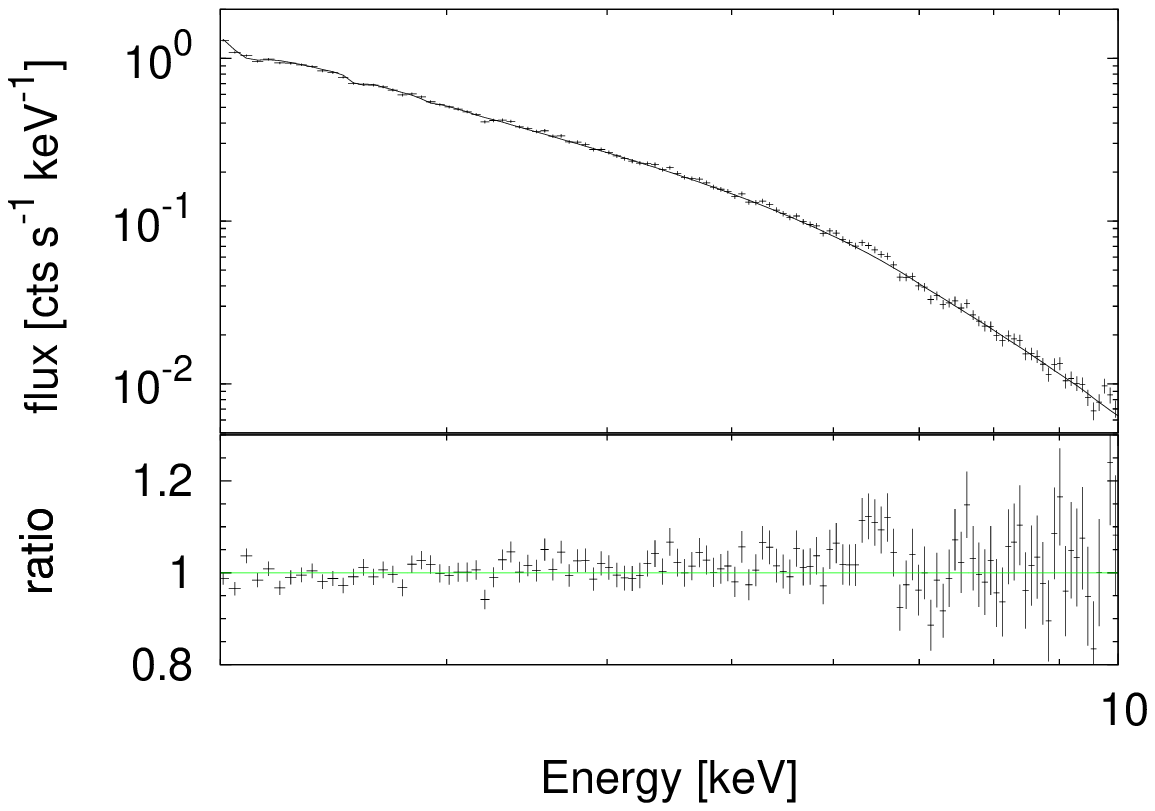} 
\includegraphics[width=0.5\textwidth]{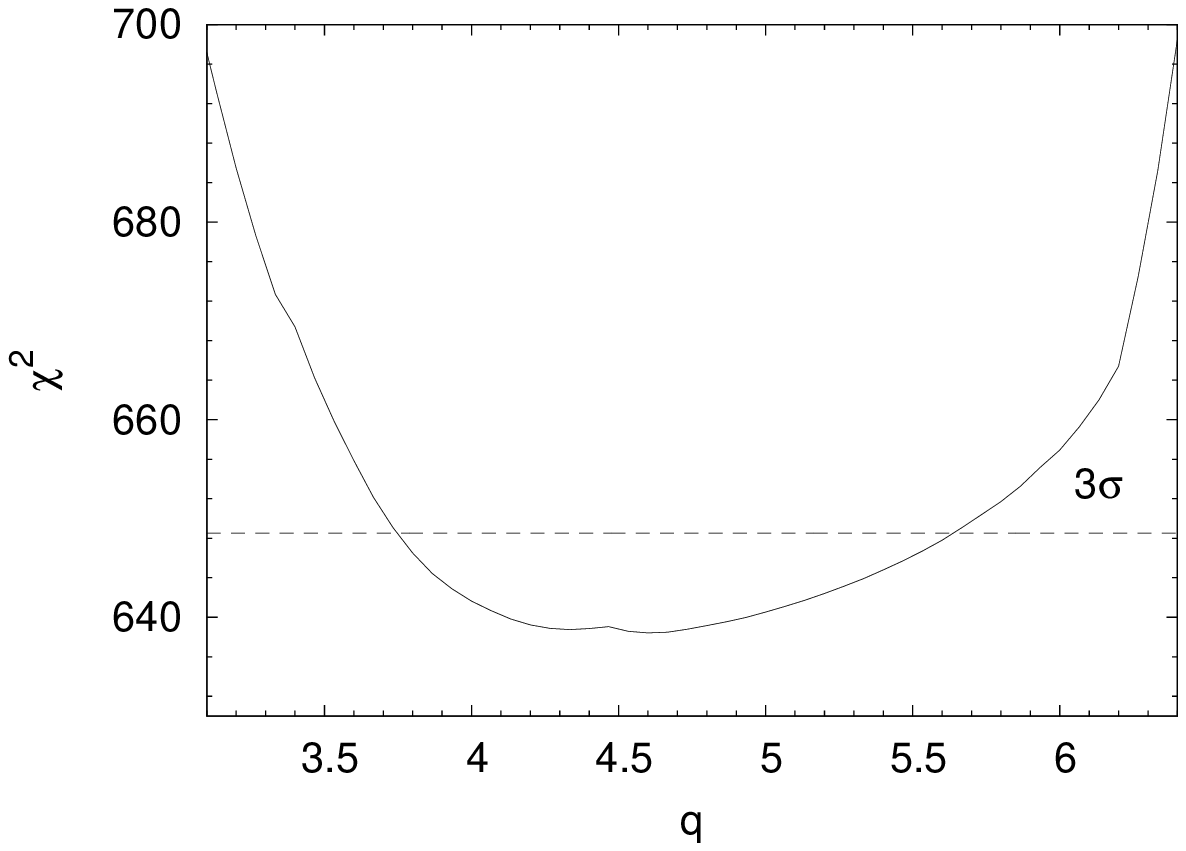}
 \caption{{\bf{Left:}} Data created by the ``complex'' model (see Figure~\ref{model_complex})
and their ratios with respect to the single-ionisation reflection model.
{\bf{Right:}} Dependence of the fit-goodness on the radial emissivity parameter of the single
reflection model. The dashed line corresponds to a $3\sigma$ confidence level.}
 \label{iso_plld_step}
\end{figure*}

\begin{figure*}
\includegraphics[width=0.5\textwidth]{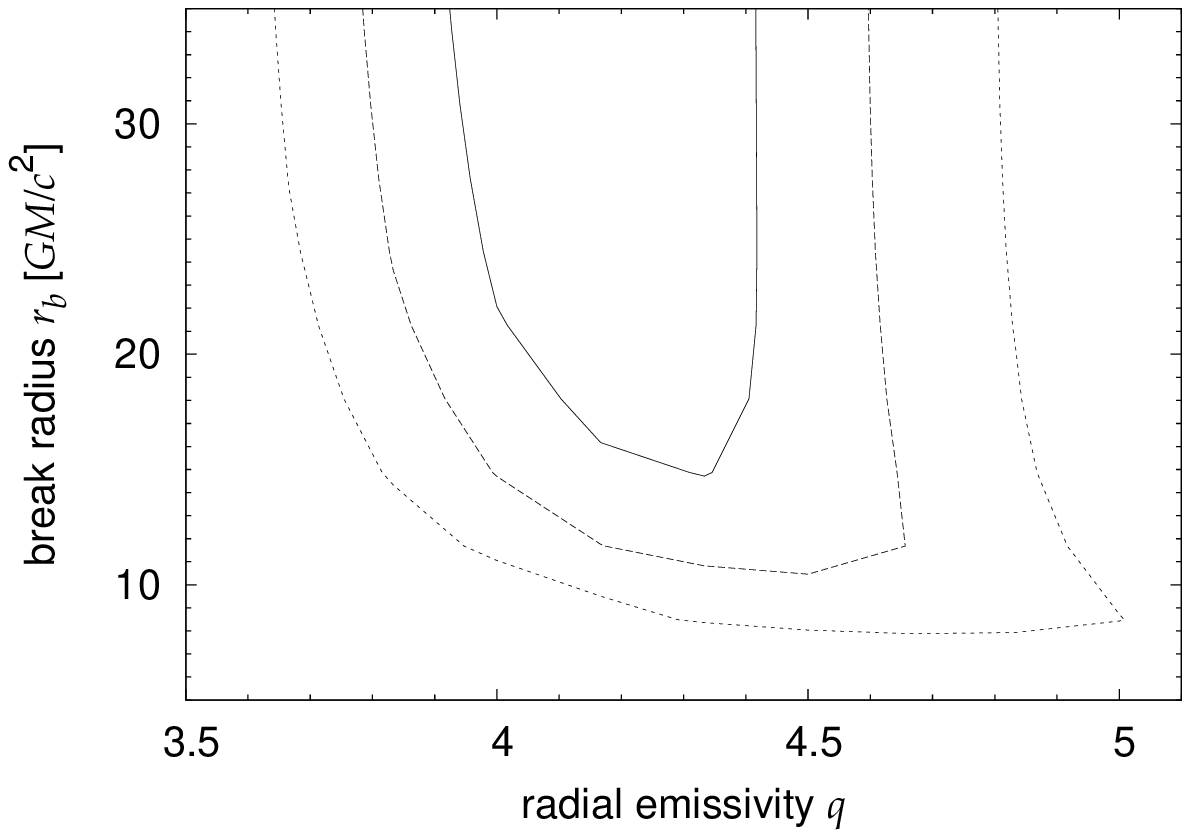}
\includegraphics[width=0.5\textwidth]{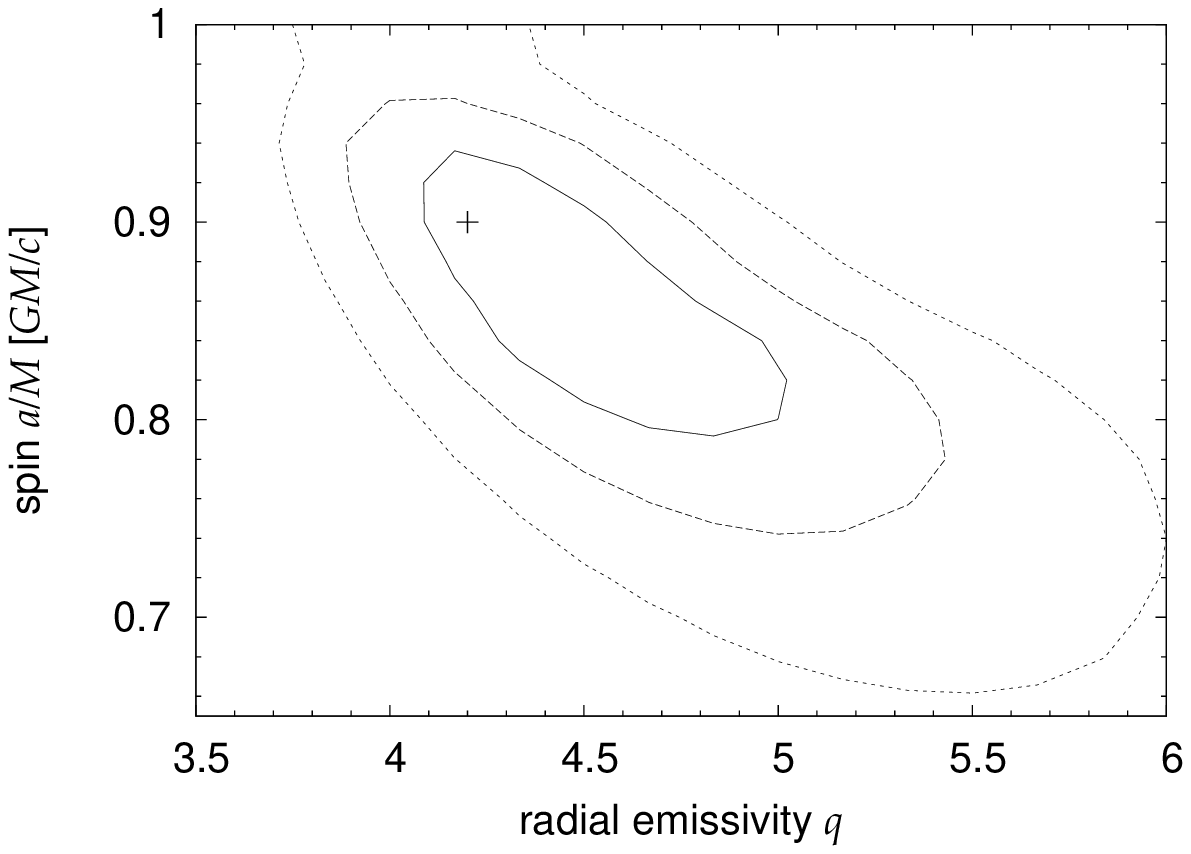}
 \caption{Contour plots of the radial emissivity parameter $q$
and the break radius $r_{\rm b}$ ({\textbf{left}}) and the spin $a$ ({\textbf{right}}),
respectively.
Data were created by a ``complex'' model
with radially stratified ionisation, 
isotropic irradiation, and angular emissivity. 
A single-ionisation model with power-law radial emissivity 
was used to fit the data.
The individual contours correspond to $1\sigma$, $2\sigma$, and $3\sigma$ level,
respectively.
}
\label{radion_iso_contours}
\end{figure*}

\begin{table}
\begin{center}
\caption{\label{table_model} Resulting parameter values of the single reflection model applied to the data simulated by a ``complex'' reflection model.}
\begin{tabular}{c|c|c} 

\hline
\hline
\rule[-0.7em]{0pt}{2em} parameter & isotropic & $h=1.5\,r_{\rm g}$ \\
\hline
\rule[-0.7em]{0pt}{2em} 
spin & $0.86^{+0.08}_{-0.07}$ & $0.94 \pm 0.02$  \\
\rule[-0.7em]{0pt}{2em} 
inner radial emissivity & $4.5^{+0.6}_{-0.5}$   & $6.7^{+0.9}_{-0.8}$ \\
\rule[-0.7em]{0pt}{2em} 
ionisation [ergs\,cm\,s$^{-1}$] & $250^{+30}_{-20}$  & $230^{+20}_{-10}$ \\
\hline
\rule[-0.7em]{0pt}{2em} 
fit goodness $\chi^2 / \nu $   & $639/606$ & $623/606$  \\
\hline
\hline
\end{tabular}
\end{center}

\tablefoot{
Two cases of assumed irradiation profile are considered,
isotropic with the radial emissivity index $q=3$,
and a point source at the height $h=1.5 r_{\rm g}$.
Default value of the spin was $a=0.94$.
}

\end{table}

We simulated a spectrum produced under these conditions
by dividing the disc into discrete rings of fixed $\xi$
and summing the spectra produced in each ring.
To generate these spectra, we used a grid of {\textsc{reflionx}} models
\citep{2005MNRAS.358..211R} with a different ionisation state convolved
with the {\textsc {KY}} model \citep{2004ApJS..153..205D}, 
corresponding to different emission regions across the disc.
For each ring, the generated reflection component
has equal integrated flux and the radial emissivity
is governed by the relativistic convolution models.
The innermost radius coincides
with the marginally stable orbit, and the outer radius was set to $400\,r_{\rm g}$.
The spin value was chosen to be $a = 0.94$, i.e. $r_{\rm ms} \approx 2\,r_{\rm g}$.
The inclination angle was chosen to be $30\,$deg. The 
photon index of the primary power-law radiation 
and its normalisation were set to
$\Gamma = 2.0$ and $K_{\Gamma}=10^{-3}$.
The standard value, $q=3$, was adopted for the radial emissivity index.
We used solar iron abundances.
%


The total reflection fraction
to the primary radiation corresponds to the case 
of an isotropically irradiated disc, i.e. similar to that of
$R \approx 1$ in the {\pexrav} model \citep{1995MNRAS.273..837M}.
Such a constructed model, which we henceforth refer to as a ``complex'' model, 
is shown with its components in Figure~\ref{model_complex}. The straight
dotted line in the plot corresponds to the power-law continuum
irradiating the disc.
The iron line around 6\,keV may be used as a diagnostic of the 
individual reflection components.
The cold reflection from the outer disc ($13.9-400\,r_{\rm g}$)
corresponds to the most prominent peak.
The second largest peak is produced by the ring occurring at $6.3-13.9\,r_{\rm g}$.
The components from the more central regions
have more smeared profile and more red-shifted iron lines.
The disc separation into individual rings was done
according to the gradient of the ionisation parameter, 
so the innermost rings are much narrower than the outer parts
and thus contribute less to the total flux
(the bottom lines in the graph).
The total reflection component is shown by a green line in the plot.

We generated the data 
in the same way as described in Sect.~\ref{data}.
We then fitted the data in the 2-10\,keV energy range with 
a model consisting of only a power-law continuum and a single 
reflection component with a broken power-law radial emissivity.
The inner radial emissivity index,
the break radius, 
the ionisation, and the normalisation of the reflection model were the only
parameters that were allowed to vary during the fitting procedure.



The resulting single-ionisation model 
represents a relatively good fit 
to the simulated data with $\chi^2 / \nu = 639/606 \approx 1.05$,
but with significant discrepancies
in the iron line and its edge.
The model (indicated by a red line) is compared with the ``complex'' reflection
in Figure~\ref{model_complex}. An apparent difference occurs
at the iron line, where a peak from distant cold reflection is missing
in the single-ionisation model.
The data residuals from this model are shown 
in the left panel of Fig.~\ref{iso_plld_step}.
The fit improved by $\Delta\chi^2 = 16$
when a Gaussian line was added to the model.
The energy and the width of the additional line were found to be
$E= 6.45^{+0.08}_{-0.07}$ and $\sigma= 0.13^{+0.07}_{-0.05}$.
The equivalent width is $41 \pm 17$\,eV.

The best-fit values of the reflection model 
parameters are summarised in the middle column of Table~\ref{table_model}.
The value of the ionisation parameter is around $250$ ergs\,cm\,s$^{-1}$.
The radial emissivity index is required to be 
significantly steeper, $q \geq 4$,
as it is also clearly visible in the right panel of Fig.~\ref{iso_plld_step}.
Figure~\ref{radion_iso_contours} (left panel) shows a contour plot between
the radial emissivity parameter and the break radius.
A preferred value of the break radius is larger than $10 r_{\rm g}$,
which demonstrates the significance of the emissivity steepening 
(i.e. that it is not related only to a narrow innermost ring).
The fit is insensitive to larger values than about $20 r_{\rm g}$.
A contour plot between the radial emissivity parameter 
and the spin is shown in the right panel of Figure~\ref{radion_iso_contours}.
A mutual degeneration of these two parameters is evident from the plot.
However, significantly steeper values of the radial emissivity
are required for any reasonable value of the spin.




\begin{figure*}
\includegraphics[width=0.49\textwidth]{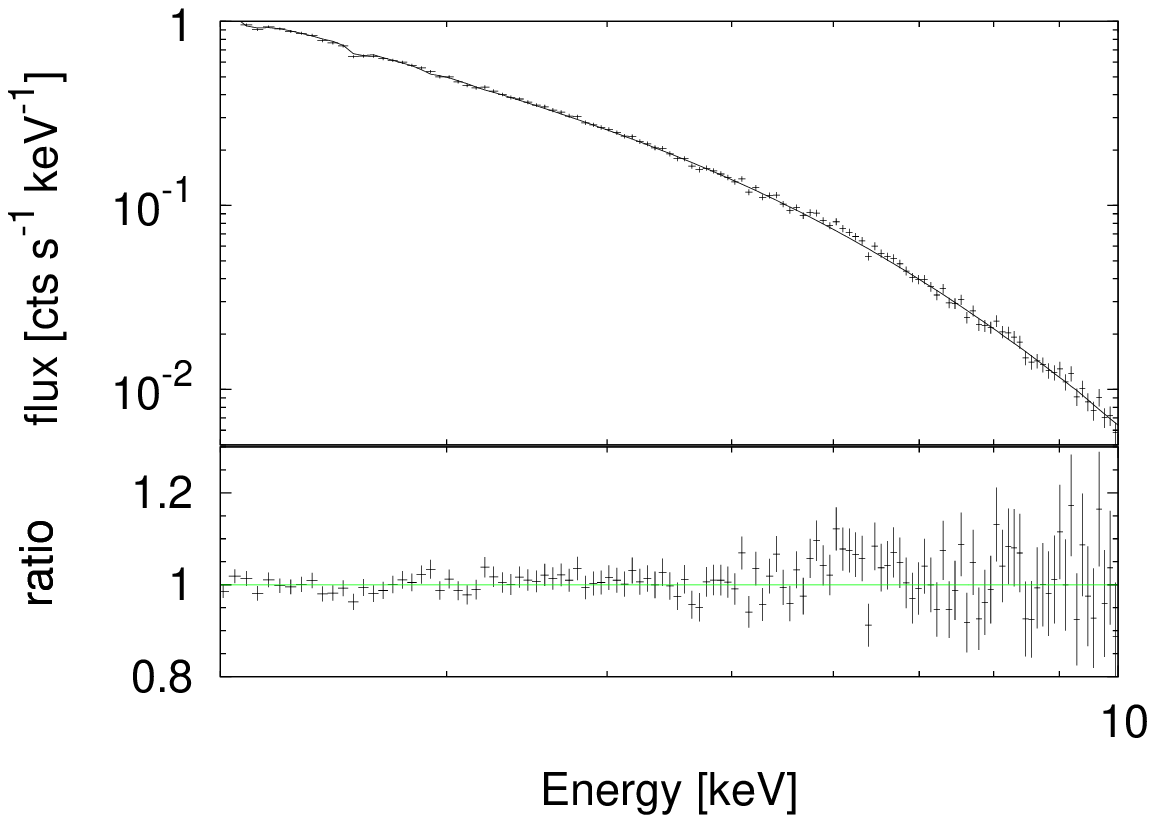}
\includegraphics[width=0.5\textwidth]{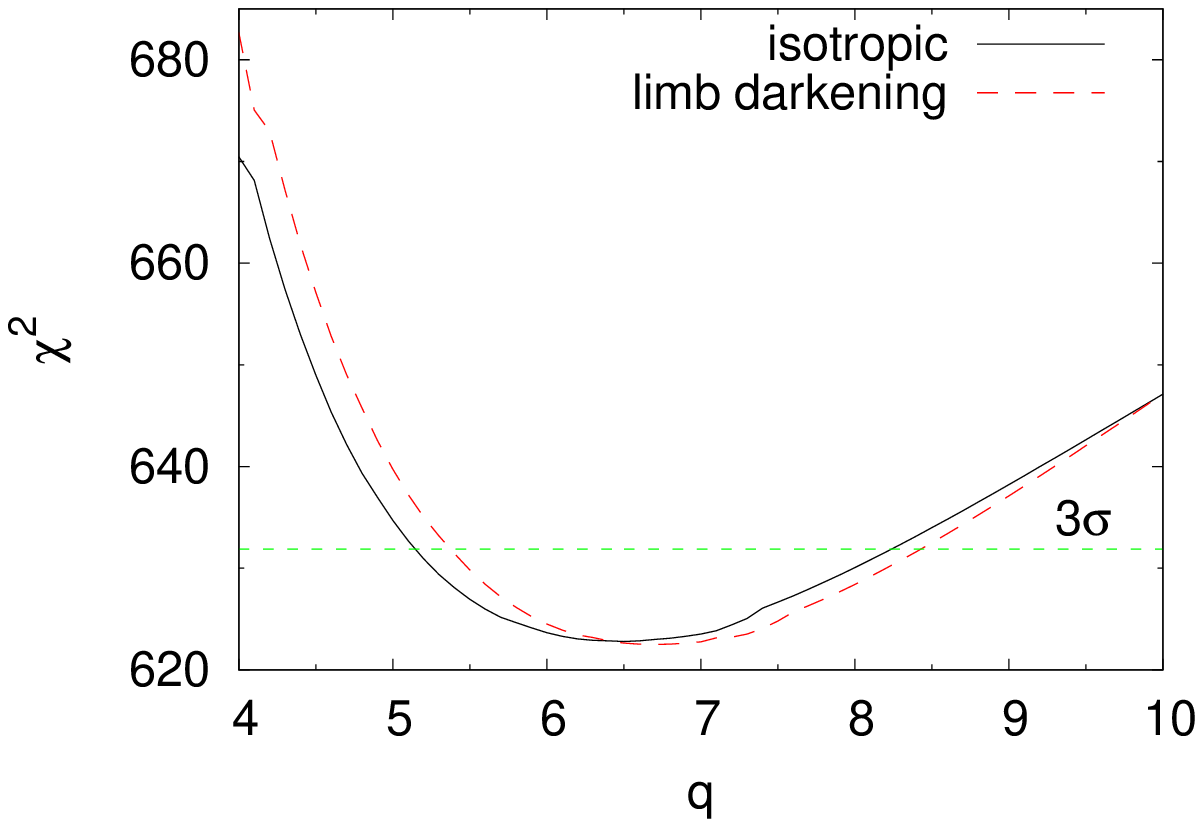}
 \caption{As in Figure~\ref{iso_plld_step}, but the data were created by a lamp-post model with the height $h=1.5\,r_{\rm g}$. In the right panel, we show 
a further overestimation of $q$ by employing limb darkening instead of isotropic
angular directionality used in the data simulation (red dashed line).}
 \label{combi_plld_step}
\end{figure*}


\section{Steep radial emissivity due to the combined effect}
\label{section_combi}

We have presented three different effects that may account for a steep radial
emissivity. However, they are not completely independent and the most likely
scenario is that they occur together.
We therefore simulated the data by considering all these
effects together. We considered a lamp-post scenario with a very low height,
$h=1.5 r_{\rm g}$. 
We calculated the ionisation profile of a constant-density disc
produced by this irradiation taking into account all the relativistic effects  
such as photon light-bending and aberration. For simplicity, we assumed an isotropic 
directionality, i.e. that the intensity of reflected radiation does not
locally depends on the emission angle. 

The data were then fitted by a single-ionisation reflection model
with the radial emissivity defined by a broken power-law with
the index $q$ as a free parameter within $30r_{\rm g}$ and $q=3$ beyond this radius.
The spectrum is reflection-dominated due to the
strong light-bending effect \citep{2004MNRAS.349.1435M}.
The parameters of the primary power-law were fixed.
The resulting model describes the data quite well with 
$\chi^2 / \nu = 623/606 \approx 1.03$,
again with some residuals in the iron-line energy band
(see the left panel of Figure~\ref{combi_plld_step}).
The best-fit values are summarised in 
the last column of Table~\ref{table_model}.

The radial emissivity is very steep, $q \approx 7$.
We show the dependence of the fit goodness on this parameter
in the right panel of Figure~\ref{combi_plld_step},
where we compare two cases of assumed directionality.
A value larger than 5 within the $3\sigma$ uncertainties
is found even with the isotropic angular emissivity.
The limb darkening is responsible for the further steepening. 
Figure~\ref{combi_contaq} shows a contour plot between the radial
emissivity and the spin. In contrast to the radial emissivity, 
the spin is found to be close to its default value.


\section{Discussion and conclusions}\label{discussion}

We have addressed the steep radial emissivities 
that have been detected in the reflection
components of the X-ray spectra of active galaxies and black-hole binaries. 
We investigated some possible explanations,
and to this end performed several simulations
to reveal the degeneracies of the radial emissivity
with other parameters and the intrinsic assumptions of the 
relativistic reflection model.
We realised that the steep radial emissivity
may be explained by either (i) the geometrical properties 
of the disc-illuminating corona, 
(ii) the use of an improper model assumption
about the angular directionality,
or (iii) radially structured ionisation.
The first puts rather extreme requirements on the corona.
It needs to be very bright
and occur at~a very low height above the black hole.
The second is likely due to use of an improper
prescription for the emission directionality in the black hole accretion disc.
The last one 
is related to a probable radial dependence 
of the disc ionisation, which plays a significant
role in the total shape of the reflection spectrum.

\subsection{Lamp-post scenario}

The steep radial emissivity may be related
to the properties of the disc-illuminating corona as suggested
before by \citet{2001MNRAS.328L..27W}.
The geometry of the emitting region certainly
plays a significant role.
A very centrally localised source at a low
height above the black hole horizon would irradiate 
the disc mainly in its central region.
The illumination in this area is greatly
enhanced by the gravitational light-bending effect
\citep[Dov\v{c}iak et al., in prep.]{2004MNRAS.349.1435M, 2008MNRAS.386..759N, 2011MNRAS.414.1269W}.

To achieve steep radial emissivity, which is assumed
to be proportional to the illumination, 
the source must be sufficiently close to the black hole.
However, in this case the primary emission has to be
extremely bright because only a small fraction would overcome 
the strong gravitational pull of the black hole and reach the 
observer \citep[see Fig.\,2 in][]{2011ApJ...731...75D}.
The importance of these effects declines sharply with height.
At heights $h \gtrsim 3\,r_{\rm g}$,
the radial emissivity profile is similar 
to the simple power-law with the standard value ($q=3$).
For even larger heights, the irradiation profile
is more complicated \citep[see Fig.\,3 and 4 in][]{2011ApJ...731...75D}.
It decreases steeply only very close to the black-hole
horizon, then becomes rather flat ($q<3$) when still in the inner parts
of the disc and finally reaches the standard value far from the centre.


\subsection{Angular directionality}

For the angular emissivity, the limb darkening law is frequently used. 
Several simulations, however, suggest that the directionality is the opposite
of limb darkening \citep[see e.g.][ and references therein]{2011A&A...527A..47R}.
The emission angle in the innermost region of the disc is always very high 
owing to strong aberration.
The flux contribution from this region is therefore underestimated
by models with limb darkening if the angular emissivity is indeed different.
This effect could lead to an approximately $20\,\%$ overestimation
of either the spin or the inner radial emissivity parameter.
\citet{2009A&A...507....1S} reanalysed the XMM-Newton observation of MCG\,-6-30-15 
and showed that the radial emissivity might be 
a more sensitive parameter to the angular
directionality than the spin. This is especially true 
when the spin value itself is very high (close to one).


\begin{figure}
\includegraphics[width=0.5\textwidth]{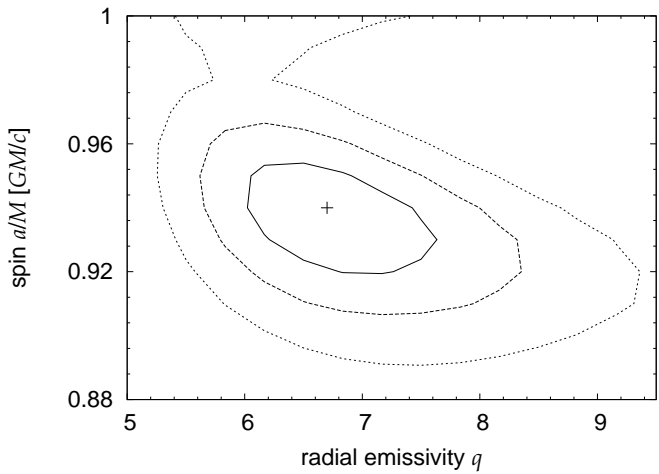}
 \caption{Contour plot of the radial emissivity parameter $q$
and the spin $a$. The data were created by a lamp-post model with 
the height $h=1.5\,r_{\rm g}$, radially stratified ionisation, 
and isotropic angular emissivity.
A single-ionisation model with limb darkening and power-law description
for the radial emissivity was used to fit the data.
The best-fit parameters are indicated by a small cross
within the contours.}
 \label{combi_contaq}
\end{figure}


\subsection{Radially structured ionisation}

We also discussed the impact of the probable radial dependence 
of the disc surface ionisation. 
The disc illumination by a corona is commonly assumed to be 
stronger in the innermost regions.
We therefore assumed that the ionisation
is higher in the innermost region and decreases with radius.
We did not consider other aspects that affect the ionisation
structure of the disc, such as the density profile,
vertical structure, and thermal processes 
(the last one being especially relevant for the stellar-mass black hole binaries).
We simply assumed that the radial dependence of the irradiation
is the dominating effect in determining the ionisation
of the disc surface. This is especially true when the lamp-post scenario
for a source of low height is considered.

We note, however, that the density may still play an important
role around the marginally stable orbit. While the Novikov-Thorne
profile of the density diverges there \citep{1973blho.conf..343N}, 
i.e. the ionisation would decrease to zero, the accretion disc may extend below
the marginally stable orbit owing to the presence of a magnetic field, 
as in the GRMHD simulations of \citet{2006ApJ...641..103H}.
The density does not increase to an infinite value, but instead
decreases close to the marginally stable orbit \citep[see also][]{2008ApJ...675.1048R}.
This region  was, for clarity, not shown in Figure~\ref{xi_profile}
but we plan to investigate it in more detail in a follow-up paper.

By fitting the simulated data,
we realised that the radial decrease in the disc ionisation 
may account for the radial-emissivity steepness 
equally as well as the assumption
of the centrally localised corona.
For the case of an isotropically illuminated disc,
we obtained radial emissivities of $q \approx 4-5$,
i.e. values similar to those for the lamp-post irradiation
of a cold disc with the height $h=1.5 r_{\rm g}$.
This is due to the different shapes of the reflection
model for different ionisation parameters.
The softer, more ionised reflection comes from the innermost part
of the disc. 
When a simplified model with a single ionisation 
is used to fit the data, it may lead to a significant
underestimation of the flux from the innermost regions,
which is then artificially compensated for by a steep value
of the radial emissivity profile in the model.

The main difference between the single-ionisation model
and the data created by a complex ionisation
occurs at the iron-line energy band and its edge 
(see Figs.~\ref{iso_plld_step} and \ref{combi_plld_step}).
While the shape of the ``complex'' reflection continuum may be 
mimicked by a~modest value of the ionisation 
and steep radial emissivity, the iron line peak
due to reflection from the cold distant parts of the disc
is not incorporated in the simplified model.
An additional emission line is then needed
to model the residuals.
In our test case, the equivalent width of such
a line was found to be about $40$\,eV.
Several AGN spectra, such as MCG\,-6-30-15 \citep{2002MNRAS.335L...1F},
have been found to contain, together with the broad disc component,
a narrow iron line associated with reflection
from a distant torus. A typical equivalent width is $\sim 100$\,eV.
The reflection from the outer parts of the accretion disc
might provide a significant contribution to this line
when the single-ionisation reflection model is applied to the disc-line modelling.

In our simulations, we considered the radial dependence of the ionisation.
However, a similar effect on the emissivity profile
may also be caused by a vertically 
structured accretion disc where the core is cold but the outer
layers are hot, i.e. strongly ionised \citep{2000ApJ...537..833N}.
\citet{2001ApJ...546..406N} showed that
in the case of AGNs, when the illuminating flux is much higher
than the thermal radiation of the disc, 
the hot skin of the disc is completely ionised 
and most of the re-processing occurs in the disc core.
However, \citet{2012MNRAS.422.1914D} revealed
the significant presence of highly ionised 
reflection together with relatively cold reflection from the disc
in the X-ray spectrum of a narrow-line Seyfert 1 galaxy 1H0707-495.
The double-reflection model is shown in their Figure~6.
Owing to their different shapes, the ionised reflection
contributes more to the red wing of the observed iron line
than the cold component. Neglecting the ionised component
would therefore result in the measurement of 
a steeper radial emissivity index.

Finally, we note that the simulations with the radially changing ionisation 
for the isotropic irradiation were done with the assumption
of a ``standard'' value for the radial emissivity, $q=3$.
However, non-thermal coronal emission does not 
necessarily need to behave in the same way 
as the thermal dissipation of the disc.
The interaction between the disc and the corona is more
complicated, including the radiation and magnetic processes
\citep[see e.g.][]{1991ApJ...380L..51H, 
2004A&A...428..353C, 2006AN....327..977G, 2011A&A...527A..47R}.
In particular, when the magnetic field is considered,
the intrinsic profile might already be as steep as $r^{-5}$
\citep{2005ApJ...635..167K}.
Further investigation of these poorly understood disc-corona interactions 
is therefore desirable.


\subsection{The combined effect}

All the effects that we have discussed are not independent of each other.
Irradiation by a compact centrally localised source is 
very anisotropic \citep{2004MNRAS.349.1435M}. 
The central regions are more illuminated and thus, 
more likely to be ionised than the outer parts of the 
accretion disc. The effect of a hypothetical
radial stratification of the disc ionisation 
is stronger than in the case of isotropic irradiation.
We therefore performed a final simulation including both
effects together. We simulated the data with the lamp-post geometry
and the ionisation stratification calculated from the theoretical
irradiation profile (assuming constant density).
We fitted the generated spectrum with a single-ionisation
model with a broken power-law for radial emissivity.
As expected, we obtained significantly steeper radial emissivity,
$q \approx 7$ (see the right panel of Table~\ref{table_model}
and Figure~\ref{combi_contaq}). 

Angular directionality plays 
an important role in the total emissivity. 
We showed that the model with limb darkening
overestimates the index of the radial emissivity profile
by $5-10\%$ (see the right panel of Figure~\ref{combi_plld_step}).
The effect might be even stronger because
our numerical calculations of local emissivity
suggest that there is a limb-brightening effect,
while we conservatively used isotropic directionality
in our simulations. Moreover, in the case
of anisotropic illumination by a lamp-post source,
the emissivity also depends on the incident
angle with a similar limb-brightening profile
(Dov\v{c}iak et al., in prep.).
The highest output of the reflected emission is thus obtained 
when both angles are grazing (perpendicular to the disc normal). 
This happens especially very close to the black hole
owing to the large aberration that is caused by the
extreme velocities of the matter in the inner disc.
Hence, the flux from the innermost region
might be significantly enhanced
and thus the radial-emissivity index overestimated.




\section{Summary}
\label{conclusions}

Very steep radial emissivity profiles with $q\sim 7$ have been previously claimed
from the analysis of the disc reflection component in the X-ray
spectra of some AGNs and black hole binaries.
We have shown here that these profiles with $q\sim 4-5$ can be a~natural consequence of either 
i) a compact centrally concentrated X-ray corona located $1-2~r_g$ away
from the central black hole, or ii) the radial dependence
of the disc ionisation state. By combining the two effects, emissivity indices as
steep as $q\sim 7$ have been obtained, which strongly suggests that objects
such as 1H~0707-495 and IRAS~13224--3809 are indeed
dominated by a centrally concentrated X-ray corona whose
radially-dependent disc irradiation induces a ionisation profile $\xi
(r) \not= {\rm{constant}}$ at the disc surface.
We have also demonstrated that an additional steepening can be simply an artifact
of improper usage of the emission directionality
in the relativistic models.

\begin{acknowledgements} 

The work was supported from
the grant ME09036 in the Czech Republic (VK).
JS and VK acknowledge support from the 
Faculty of the European Space Astronomy Centre (ESAC).
GM thanks the Spanish Ministerio de
Ciencia e Innovaci\'on and CSIC for support through a Ram\'on y Cajal
contract. Financial support for this work was also provided by the
Spanish Ministry of Science and Innovation (now Ministerio de
Econom\'a y Competitividad) through grant AYA2010-21490-C02-02.

\end{acknowledgements}

\bibliographystyle{aa} 
\bibliography{svoboda} 

\end{document}